\documentclass[amsmath,amssymb,aps,reprint]{revtex4-1}

\usepackage{graphicx}
\usepackage{dcolumn}
\usepackage{bm}

\usepackage[english]{babel}
\usepackage[T1]{fontenc}
\usepackage[utf8]{inputenc}
\usepackage[thinspace,mediumqspace]{SIunits}
\newcommand{\tens}[1]{\overset{\leftrightarrow}{#1}}
\newcommand{\guil}[1]{``{#1}''}
\begin{document}

\preprint{APS/123-QED}

\title{Hall-effect sign-inversion in a realizable 3D metamaterial}

\author{Muamer Kadic$^1$}
\author{Robert Schittny$^1$}
\author{Tiemo B\"uckmann$^1$}
\author{Christian Kern$^1$}
\author{Martin Wegener$^{1,2}$}
\affiliation{$^1$Institute of Applied Physics, Karlsruhe Institute of Technology (KIT), 76128 Karlsruhe, Germany}%
\affiliation{$^2$Institute of Nanotechnology, Karlsruhe Institute of Technology (KIT), 76344 Eggenstein-Leopoldshafen, Germany}


\date{\today}

\begin{abstract}
In 2009, Briane and Milton proved mathematically the existence of three-dimensional isotropic metamaterials with a classical Hall coefficient which is negative with respect to that of all of the metamaterial constituents. Here, we significantly simplify their blueprint towards an architecture composed of only a single constituent material in vacuum/air, which can be seen as a special type of porosity. We show that the sign of the Hall voltage is determined by a separation parameter between adjacent tori. This qualitative behavior is robust even for only a small number of metamaterial unit cells. The combination of simplification and robustness brings experimental verifications of this striking sign-inversion into reach.
\end{abstract}

\keywords{Hall effect, metamaterial, negative properties}
   
\maketitle

\nocite{*}

It is often stated that measuring the sign of the Hall voltage allows for determining whether negatively charged electrons or positively charged holes dominate the electrical transport in ordinary materials such as semiconductors or metals \cite{Hall1879,Jackson1998,Milton2002}. In 2009, Briane and Milton \cite{Brian2009} proved mathematically that this statement does not apply to arbitrary artificial materials: They considered a three-dimensional simple-cubic arrangement of interlinked nearly touching tori made of a first conductive material, a second material in between adjacent tori, and a third conductive embedding material (see Fig.\,1). They showed that the sign of the resulting effective Hall coefficient $R_{\rm H}^{\rm eff}$ can be opposite to that of the Hall coefficients of the three constituents $R_{\rm H}$, i.e., $R_{\rm H}^{\rm eff} \cdot R_{\rm H}<0$. Furthermore, they proved analytically that the behavior of the structure in Fig.\,1 is isotropic and that a corresponding sign-inversion is not possible in two dimensions \cite{Brian2009,Brian2012}.

\begin{figure}[ht!]
	\includegraphics[width=0.48\textwidth]{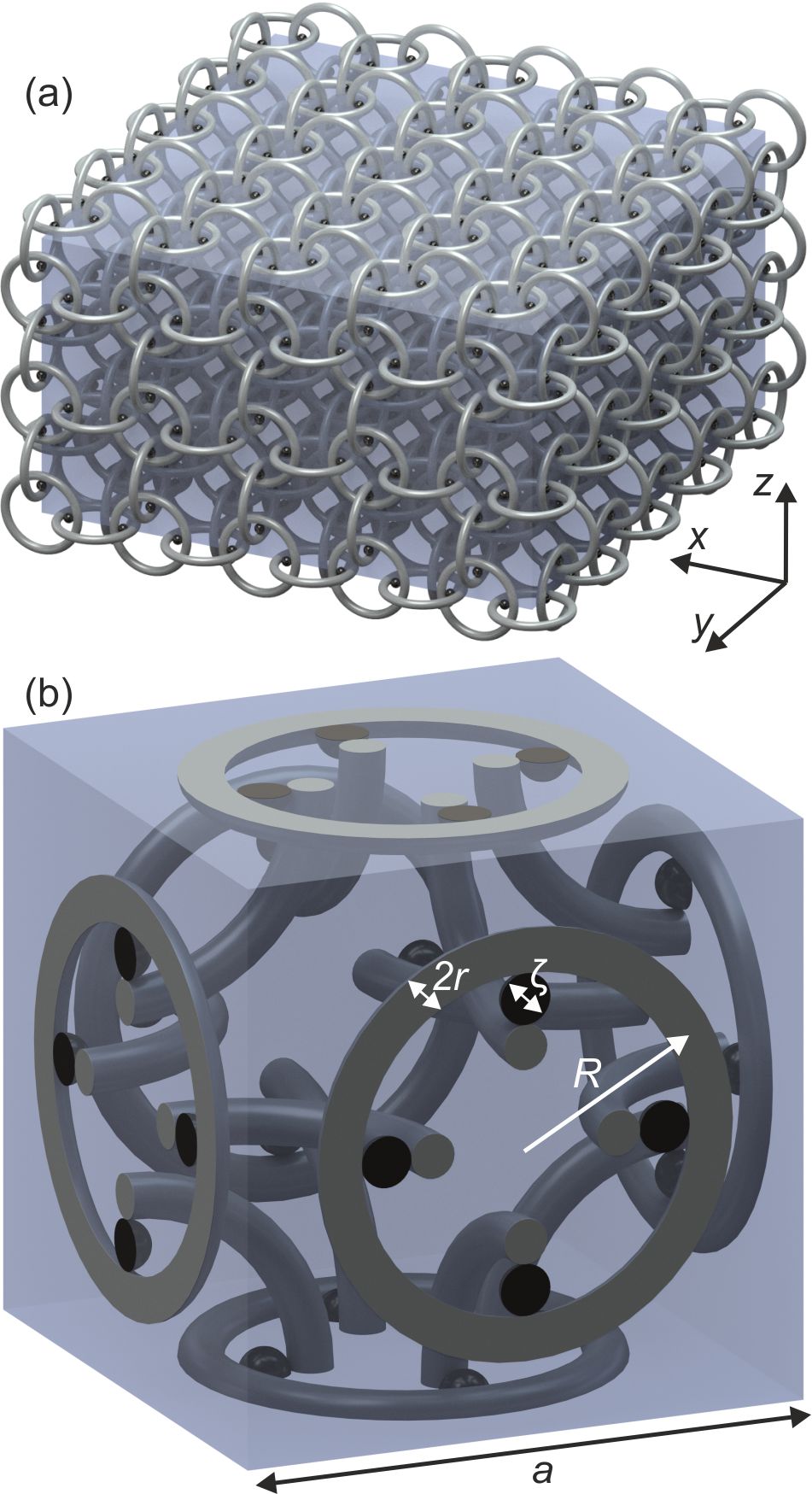}%
	\caption{(a) Crystalline three-dimensional simple-cubic metamaterial structure proposed by Briane and Milton \cite{Brian2009}. (b) One cubic unit cell with lattice constant $a$. It is composed of three different isotropic constituent materials (the tori, the spheres, and the surrounding), all of which have the same sign of the Hall coefficient, yet different magnitudes. Briane and Milton showed analytically that the resulting effective metamaterial Hall coefficient is isotropic and can have the opposite sign.}
	\label{fig1}
\end{figure}

Their finding \cite{Brian2009} emphasizes that the effective properties of artificial materials called metamaterials need not be intermediate to the properties of their constituents. Related examples have served as a catalyzer for the emerging field of metamaterials: At some finite angular frequency $\omega>0$, the effective electric permittivity $\epsilon_{\rm eff} (\omega)$ of a composite \cite{Bergman1978,Bergman1983,Milton2002} made of, e.g., one isotropic material with $\epsilon (\omega)=-1$ in vacuum (with $\epsilon_{\rm vac}(\omega)=1$), can assume any value $\epsilon_{\rm eff}(\omega) \in [-\infty,\infty]$. Likewise, a metamaterial composed of one non-magnetic isotropic constituent with magnetic permeability $\mu(\omega)=1$ in vacuum (with $\mu_{\rm vac}(\omega)=1$), allows for $\mu_{\rm eff} (\omega) \in [-\infty,\infty]$. By combination of these two examples, composites of constituent materials with (real part of the) refractive index $n(\omega) \geq 0$ can lead to $n_{\rm eff} (\omega) \cdot n(\omega)<0$ \cite{Pendry1999,Milton2002,Soukoulis2007}. However, for electric current conduction, the possibility of such opposite sign has not been known until the work of Briane and Milton \cite{Brian2009}. Furthermore, their example concerns the static case, $\omega=0$, whereas the other mentioned examples are inherently restricted to finite frequencies $\omega \neq 0$.

In this paper, we start by verifying the analytical Briane-Milton result using numerical calculations, we then drastically simplify their suggestion to a single porous material, we show that the sign and magnitude of the Hall coefficient can be changed by the geometry of the pores, and we demonstrate that the behavior converges for an accessible number of metamaterial unit cells. On this basis, experimental realizations come into reach.

\begin{figure}[ht!]
	\includegraphics[width=0.48\textwidth]{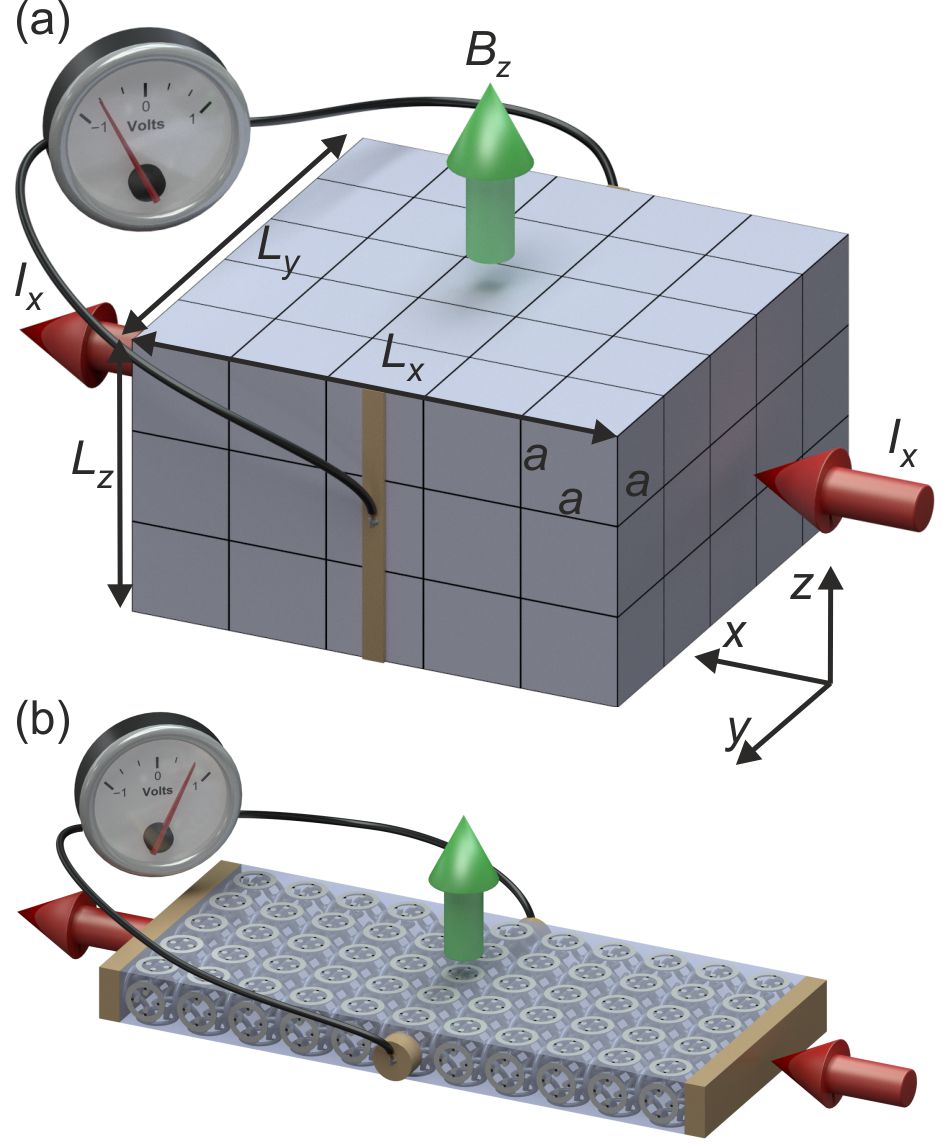}%
	\caption{(a) Scheme of a cuboid isotropic homogeneous material exhibiting the classical Hall effect with Hall voltage $U_{\rm H}$, resulting from a constant electric current $I_x$ and a perpendicular static magnetic field $B_z$. The shown geometry and coordinate system is used throughout this paper. The small cubes with side length $a$ indicate the unit cells of the crystalline material with integer numbers of unit cells $N_x$, $N_y$, and $N_z$ in the $x$-, $y$-, and $z$-direction. This leads to the sample dimensions $L_x=N_x\,a$, $L_y=N_y\,a$, and $L_z=N_z\,a$. (b) Hall bar composed of $N_x=11$, $N_y=5$, and $N_z=1$ metamaterial unit cells like shown in Fig.\,1(b). The Hall bar contains four metal contacts ($=$ equi-potential surfaces) to impose the current $I_x$ and to pick up the Hall voltage $U_{\rm H}$, respectively.}
	\label{fig2}
\end{figure}

In the textbook version of the classical Hall effect illustrated in Fig.\,2(a), a constant electrical current $I_x$ (corresponding to an applied voltage $U_x$) is drawn along the $x$-direction through a cuboid isotropic ordinary material with volume $L_x L_y L_z$. A static magnetic field $B_z$ is applied along the $z$-direction. We will use this geometry throughout this paper. The Lorentz force leads to a potential difference along the $y$-direction called the Hall voltage, which is given by $U_{\rm H}=R_{\rm H} I_x B_z/L_z$, with the Hall coefficient $R_{\rm H}=1/\rho$. The electric-charge density $\rho$ can be positive or negative. Ideally, the Hall voltage does not depend on the extent of the sample in the $x$- and $y$-directions, $L_x$ and $L_y$, and scales inversely with the sample thickness $L_z$. Thus, one usually considers thin films to obtain large Hall voltages, or even a single atomic layer in the case of the quantum Hall effect on graphene \cite{Zhang2005}. Fig.\,2(b) illustrates a Hall-bar geometry composed of a total of $11\times5\times 1=55$ metamaterial unit cells as to be used in several of the calculations to be presented below.

\begin{figure}[h!]
	\includegraphics[width=0.45\textwidth]{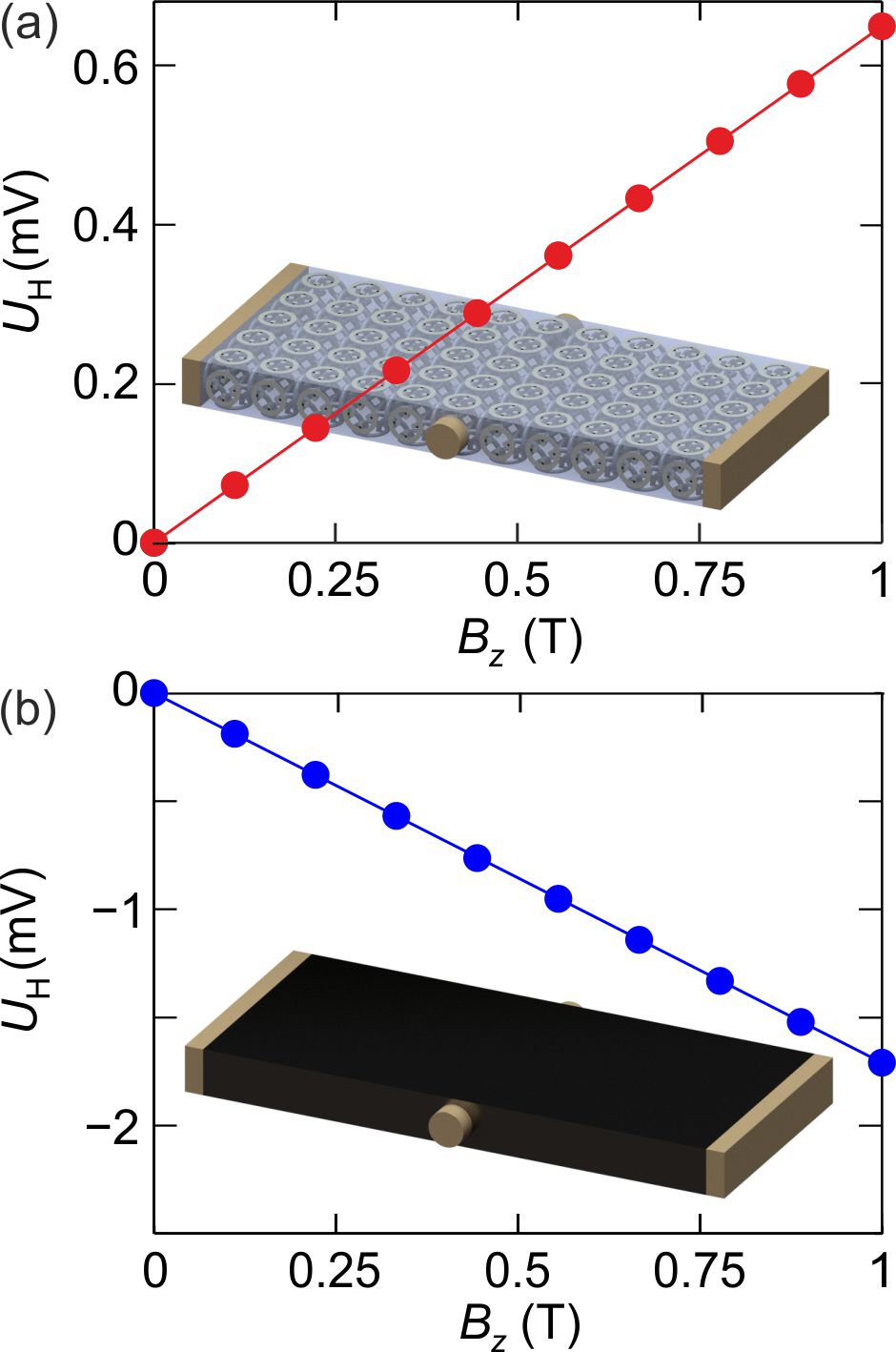}%
	\caption{(a) Calculated Hall voltage $U_{\rm H}$ (dots) for the Briane-Milton metamaterial shown in Fig.\,1 for the geometry in Fig.\,2(b) versus applied magnetic field $B_z$ for $N_x=11$, $N_y=5$, and $N_z=1$, i.e., for a total of $55$ metamaterial unit cells. The straight line is a guide to the eye. Each constituent alone would show a negative Hall voltage in bulk form. Material parameters for the spheres $R_{\rm H}=- 624 \times 10^{-6}\; {\rm m^3 A^{-1} s^{-1}}$, $\sigma_0=200 \; {\rm AV^{-1} m^{-1}}$, for the tori $R_{\rm H}=- 3.5 \times 10^{-11}\; {\rm m^3 A^{-1} s^{-1}}$, $\sigma_0=3.5 \times 10^7 \; {\rm AV^{-1} m^{-1}}$, and for the surrounding $R_{\rm H}=0 \, {\rm m^3 A^{-1} s^{-1}}$ and $\sigma_0=2 \; {\rm AV^{-1} m^{-1}}$. The current is $I_x=0.1 \, {\rm mA}$, the geometrical parameters are $R=10\, {\rm \mu m}$, $r=1.6\, {\rm \mu m} $, $\zeta=1.6 \, {\rm \mu m}$ (for comparison, $\zeta=2r$ in Fig.\,1), hence $a=4R-2(\zeta+2r)=30.4\,\rm \mu m$. The cylindrical pick-up contacts have radius $R+r$, height $0.2 \times R$, and  conductivity $3.5 \times 10^7 \; {\rm AV^{-1} m^{-1}}$ (aluminum). (b) Calculated Hall voltage (dots) for a bulk material with the same constituent-material parameters as the spheres in panel (a). Note the different vertical scales. The straight line is a guide to the eye.}
	\label{fig3}
\end{figure}

The magnetic component of the Lorentz force makes the configuration effectively anisotropic. A more detailed mathematical description thus has to start from the static version of the continuity equation for the electric current density vector $\vec{j}$ given by
\begin{equation}
\vec{\nabla} \cdot \vec{j}=\vec{\nabla} \cdot (\tens{\sigma} \vec{\nabla} \phi)=0 \;\,
\end{equation}
with the scalar electrostatic potential $\phi= \phi (\vec{r})$ and the (anisotropic) electric conductivity tensor \cite{Milton2002}
\begin{equation}
\tens{\sigma}= \begin{pmatrix} \frac{\sigma_0}{1+(\sigma_0 R_{\rm H} B_z )^2} & \frac{\sigma_0^2 R_{\rm H} B_z }{1+(\sigma_0 R_{\rm H} B_z )^2} & 0\\ -\frac{\sigma_0^2 R_{\rm H} B_z}{1+(\sigma_0 R_{\rm H} B_z )^2} & \frac{\sigma_0}{1+(\sigma_0 R_{\rm H} B_z )^2} & 0\\ 0 & 0 & \sigma_0\\ \end{pmatrix}\,.
\end{equation}

\begin{figure}[h!]
	\includegraphics[width=0.4\textwidth]{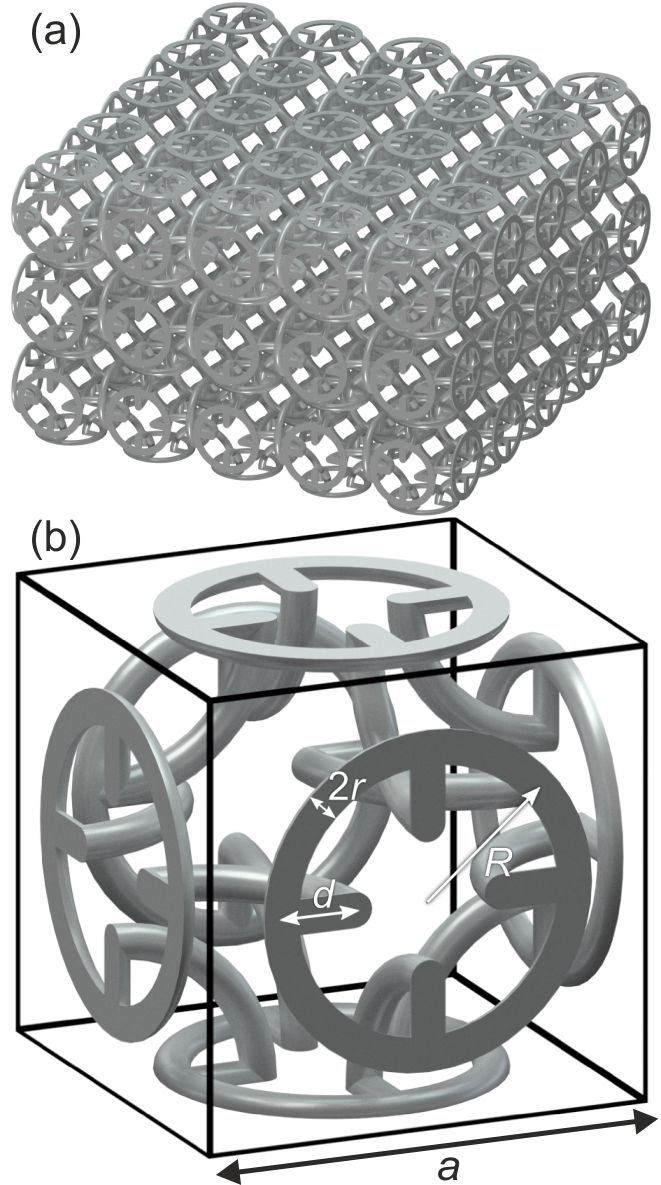}%
	\caption{(a) Simplified three-dimensional metamaterial (compare Fig.\,1) composed of a single constituent material only. We take n-doped silicon with conductivity $\sigma_0=200 \; {\rm AV^{-1} m^{-1}}$ and Hall coefficient $R_{\rm H}=-624 \times 10^{-6}\; {\rm m^3 A^{-1} s^{-1}}$. (b) One cubic unit cell with lattice constant $a$. The porous structure can lead to a Hall voltage with inverted sign with respect to that of the homogeneous bulk material (see Fig.\,5). A crucial parameter is the separation $d$ between adjacent tori, which can be positive, zero, or negative (see Fig.\,5). The depicted configuration corresponds to $d<0$. The torus diameter is $2R$ and the torus wire diameter $2r$. In this paper, we fix $R=10\,\rm \mu m$. The lattice constant $a$ results from $a=4R+2d$.}
	\label{fig4}
\end{figure} 

Here, $\sigma_0$ is the scalar conductivity of the isotropic material for $B_z=0$. For small magnetic fields, i.e., in the limit $(\sigma_0 R_{\rm H} B_z )^2 \ll 1$, all denominators can be approximated by $1$ and we get the usual proportionality $U_{\rm H} \propto R_{\rm H} B_z$. In this tensor form, the mathematics can be applied to spatially inhomogeneous materials as well, for which $\tens{\sigma} \rightarrow \tens{\sigma} (\vec{r}) $, with $\sigma \rightarrow \sigma (\vec{r})$ and $\rho \rightarrow \rho (\vec{r})$, and hence $R_{\rm H} \rightarrow R_{\rm H} (\vec{r})$ (while $B_z={\rm const.}$).

In this paper, in all calculations for just a single constituent material, we will use constituent material parameters typical for n-doped silicon (e.g., doped with phosphorus) at room temperature \cite{Sze1985}, namely $\sigma_0=200 \, {\rm AV^{-1} m^{-1}}$ and a doping density (electron density) of $n_{\rm e}=10^{22}\, {\rm   m^{-3}}$. This choice leads to a charge density of $\rho=-{\rm e} n_{\rm e}$, with the electron charge $-{\rm e}=-1.602 \times 10^{-19}\,\rm As$, and to a Hall coefficient of $R_{\rm H}=1/\rho =- 624 \times 10^{-6}\, {\rm m^3 A^{-1} s^{-1}}$. These values are not critical at all and the results presented below can, in fact, easily be scaled to any other parameter combination.

\begin{figure}[h!]
	\includegraphics[width=0.45\textwidth]{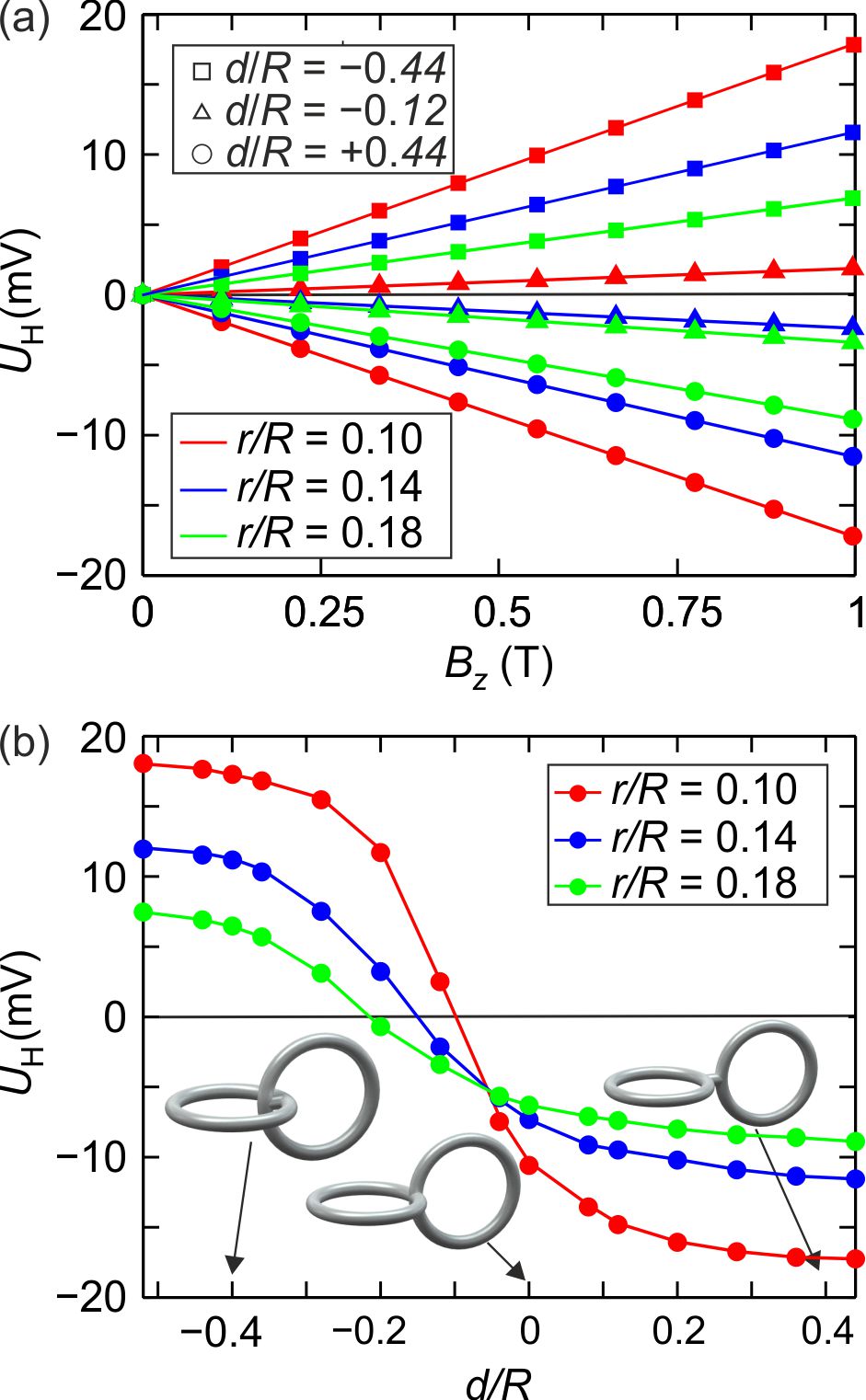}%
	\caption{(a) Calculated Hall voltage $U_{\rm H}$ (dots) for the simplified single-component metamaterial shown in Fig.\,4 for the geometry in Fig.\,2 and for fixed $I_x=0.1\,\rm mA$ versus applied magnetic field $B_z$ for various parameters. The straight lines are guides to the eye. In bulk form, the constituent material would lead to a negative Hall voltage. (b) Calculated Hall voltage (dots) versus separation parameter $d$ (see Fig.\,4) divided by the torus radius $R$. The curves are guides to the eye. Obviously, the sign of the Hall voltage correlates with the ratio $d/R$, whereas the wire radius $r$ has a much smaller influence. The insets illustrate the geometry for three different values of $d/R$. Parameters are: $R=10\,\rm \mu m$, $B_z=1\,\rm T$, $N_x=11$, $N_y=5$, $N_z=1$, contacts as shown in Fig.\,2(b), and n-doped silicon with $\sigma_0=200 \, {\rm AV^{-1} m^{-1}}$ and $R_{\rm H}=-624 \times 10^{-6}\, {\rm m^3 A^{-1} s^{-1}}$.}
	\label{fig5}
\end{figure}

\begin{figure}[h!]
	\includegraphics[width=0.48\textwidth]{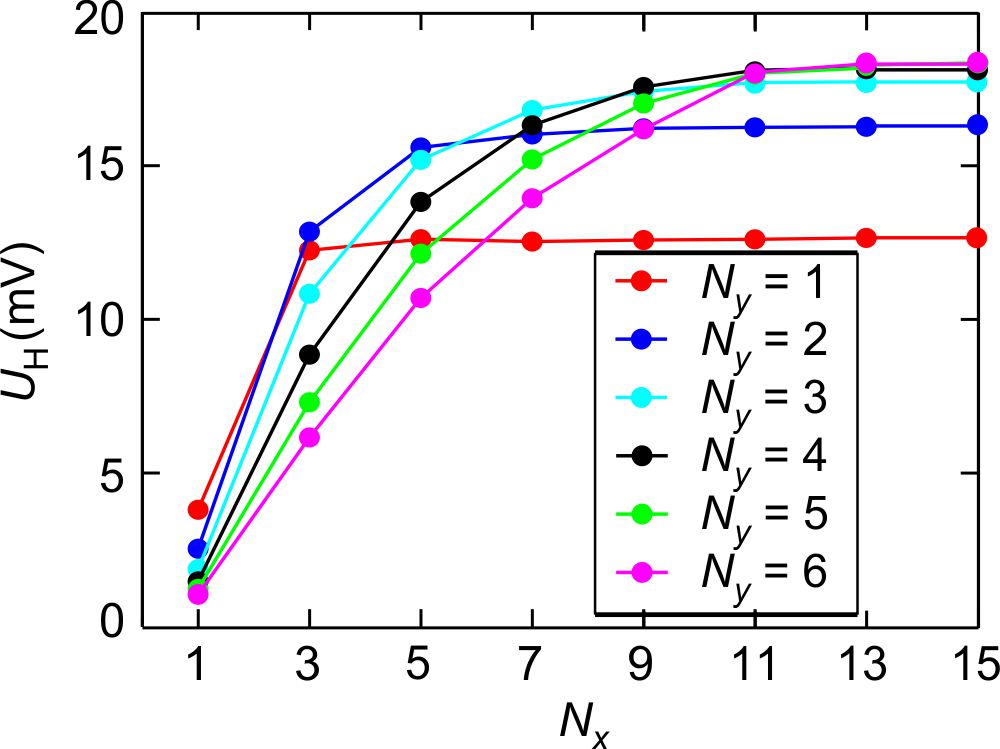}%
	\caption{Calculated Hall voltage $U_{\rm H}$ (dots) for fixed geometry and parameters of the metamaterial unit cell versus $N_x$ for different values of $N_y$ as indicated in the legend. We fix $N_z=1$. The shape of the contacts is as shown in Fig.\,2(b). The other fixed parameters are $\sigma_0=200 \, {\rm AV^{-1} m^{-1}}$, $R_{\rm H}=-624 \times 10^{-6}\, {\rm m^3 A^{-1} s^{-1}}$, $R=10\,\rm \mu m$, $r/R=0.1$, $d/R=-0.44$, $B_z=1\,\rm T$ and $I_x=0.1\,\rm mA$}
	\label{fig6}
\end{figure}

\begin{figure}[h!]
	\includegraphics[width=0.48\textwidth]{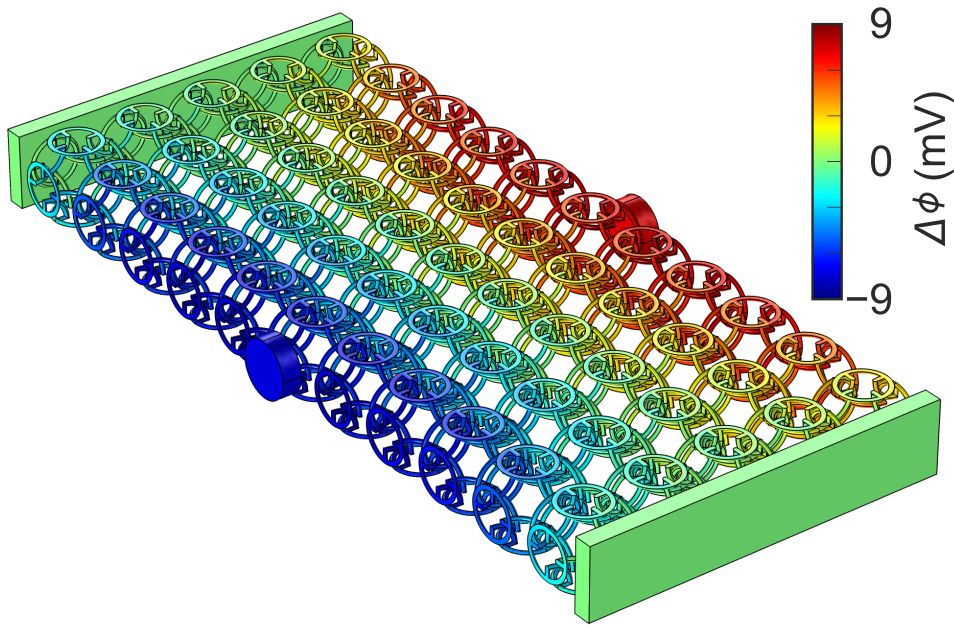}
	\caption{Hall potential map, i.e., difference of the calculated electrostatic potential $\phi(\vec{r})$ with ($B_z=1\,\rm T$) and without ($B_z=0$) magnetic field for the parameters as in Fig.\,6 for the case of $N_x=11$, $N_y=5$, and $N_z=1$ plotted on a false-color scale.}
	\label{fig7}
\end{figure}

Our numerical calculations have been performed using the software package COMSOL Multiphysics (Mumps solver). Typically, we discretize the entire Hall bar composed of the metamaterial unit cells and the four metallic contacts (= \; equi-potential surfaces) by $\approx 2 \times 10^6$ tetrahedral elements. All results depicted in this paper are in the limit of sufficiently small magnetic fields, for which the Hall voltage $U_{\rm H}$ is proportional to the external magnetic field $B_z$. We intentionally avoid, however, considering too small magnetic fields for which the off-diagonal elements of the above conductivity tensor are many orders of magnitude smaller than the diagonal elements, leading to numerical difficulties. Furthermore, the meshing is the same for each unit cell and  symmetric with respect to the $y$-direction. Otherwise, artificial transverse \guil{Hall} voltages can occur for zero magnetic field.

We start our discussion by considering the geometry and the parameters chosen by Briane and Milton \cite{Brian2009} (see Fig.\,1) with a Hall bar composed of $55$ metamaterial unit cells like shown in Fig.\,2(b). In Fig.\,1(b), the spheres touch the tori at singular points only. This configuration is pathological numerically. We have thus rather considered a small but finite contact area. Corresponding numerical results are depicted in Fig.\,3(a). They reproduce the analytically predicted sign-inversion of the Hall voltage with respect to the bulk material shown in Fig.\,3(b). 

Such intricate three-dimensional microstructures containing three different constituents can possibly be realized experimentally, but their microfabrication is quite demanding by today's standards. A central aspect of this paper is thus to simplify the microstructures as much as possible, which is possible indeed: We obtain similar results for using only a single constituent material in vacuum/air (illustrated in Fig.\,4, n-Si parameters as quoted above). For example, for a moderate magnetic field of $B_z=1 \; \rm T$ and a constant current of $I_x=0.1\; \rm mA$, we calculate a Hall voltage in the range of $10\,\rm mV$  as shown in Fig.\,5(a). The combination of these parameters appears quite reasonable for practical experimental realizations.

Figure 5(b) depicts results for different separations of adjacent tori, $d$ (as defined in Fig.\,4), while fixing the constituent material and while fixing the radius of the tori to $R=10\,\rm \mu m$. As a result, the metamaterial lattice constant $a$ changes according to $a=4R+2d$ (compare Fig.\,4). Obviously, the sign of the Hall voltage changes versus $d$: For thin tori wires, $r/R\ll 1$ (see Fig.\,4), the zero crossing of the Hall voltage $U_{\rm H} (d)$ asymptotically occurs at $d=0$, whereas it is somewhat shifted from zero for larger ratios $r/R$.  

\begin{figure}[h!]
	\includegraphics[width=0.48\textwidth]{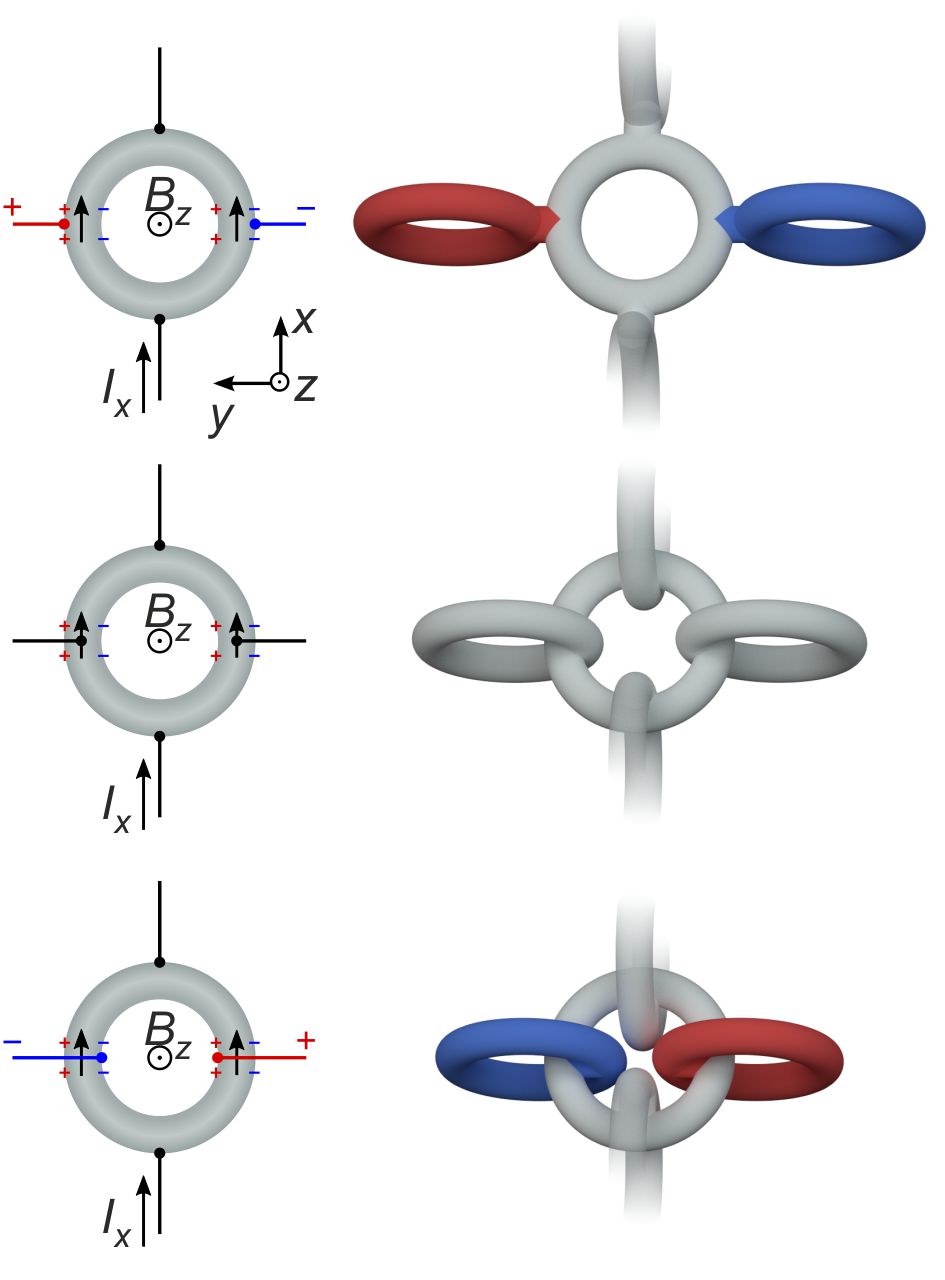}
	\caption{Illustration of the sign-inversion of the Hall voltage for different tori separations $d$. From top to bottom: $d>0$, $d\approx 0$, and $d<0$. The schemes in the left-hand side column can be compared with parts of the metamaterial structure shown in the right-hand side column.}
	\label{fig8}
\end{figure}

For bulk Hall media, it is well known \cite{Pauw1958,Moelter1998,Popovic2003} that the actual Hall voltage depends on the geometry. In case of cuboid geometry like shown in Fig.\,2, the thickness $L_z$ is usually chosen to be small compared to the other dimensions to maximize the Hall voltage. Furthermore, the width $L_y$ has to be small compared to the length $L_x$ and the pick-up contacts have to be small compared to the width in order to minimize the influence of the contacts.  In addition, for metamaterials, unlike for ordinary crystals, the unit cell is no longer very small compared to the size of the Hall bar. We have thus investigated the dependence of the Hall voltage on the number of metamaterial unit cells for fixed shape and parameters of the unit cell. Results are depicted in Fig.\,6. Altogether, these results suggest that a Hall bar composed of $N_x=11$, $N_y=5$, and $N_z=1$ metamaterial unit cells, under the quoted conditions, together with the contacts shown in Fig.\,2(b) approximates reasonably well a fictitious infinitely extended Hall-effect continuum. If the pick-up contacts (=\, equi-potential surfaces) are artificially removed, the Hall voltage changes by a very few percent with respect to the case with pick-up contacts (Fig.\,5).

In Fig.\,7, we depict the difference of the electrostatic potential $\phi$ with ($B_z=1\,\rm T$) and without ($B_z=0$) magnetic field for the parameters as in Fig.\,6 for the case of $N_x=11$, $N_y=5$, and $N_z=1$. Obviously, the edge effects decay towards the center of the Hall bar, again indicating that we approach the conceptual ideal of an infinitely extended Hall-effect continuum with a finite and potentially accessible number of metamaterial unit cells.  

\begin{figure}[h!]
	\includegraphics[width=0.45\textwidth]{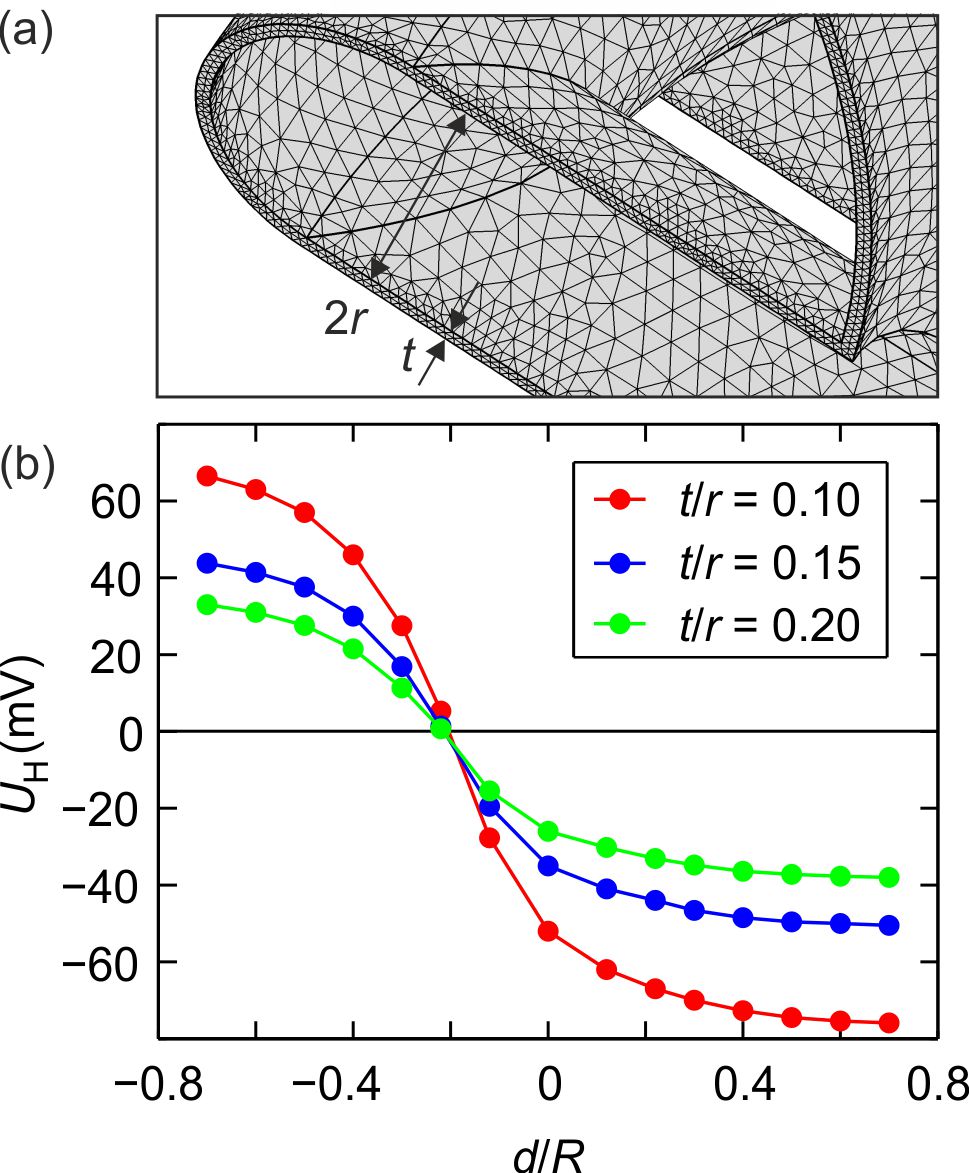}%
	\caption{(a) Hollow version of the single-component metamaterial in Fig.\,4. The shown mesh has been used for the numerical calculations. (b) Calculated Hall voltage (dots) as Fig.\,5(b) but for the hollow structure shown in panel (a). Parameters are: $N_x=11$, $N_y=5$, $N_z=1$, contacts as shown in Fig.\,2(b), $I_x=0.1 \,{\rm mA}$, $B_z=1\,\rm T$, $R=10\,\rm \mu m$, $r/R=0.14$, and n-doped silicon with $\sigma_0=200 \, {\rm AV^{-1} m^{-1}}$ and $R_{\rm H}=-624 \times 10^{-6}\, {\rm m^3 A^{-1} s^{-1}}$. The three different coating thicknesses $t/r$ are indicated in the legend.}
	\label{fig9}
\end{figure}

The dependence $U_{\rm H}(d)$ shown in Fig.\,5(b) is the key to an intuitive understanding of the Hall-effect sign-inversion in the single-constituent metamaterial. To start with, note that the tori parallel to the $yz$-plane lead to negligibly small local Hall voltages along the $y$-direction because the vector product of current and magnetic field has no $y$-component for these tori. The situation is different for the tori parallel to the $xy$-plane and those parallel to the $xz$-plane. 

Consider the tori parallel to the $xy$-plane illustrated in Fig.\,8. At their center along $x$, the current essentially flows in the $x$-direction. Together with the magnetic field parallel to $z$, this current leads to an ordinary Hall voltage along the $y$-direction across the torus wire thickness on the left as well as on the right-hand side of the torus. For $d>0$, the tori parallel to the $yz$-plane essentially pick up this voltage. In contrast, for $d\approx 0$, these tori connect to points of equivalent potential and the external Hall voltage is zero. For $d<0$, the $yz$-torus on the left of the $xy$-torus connects to the right-hand side of the torus wire and vice versa. As a result, the $yz$-tori pick up the local Hall voltage and guide it to the outside with an inverted sign. This reasoning implies that the separation $d_x$ along the $x$-direction is of minor importance---although the current locally flows in the negative $x$-direction at these connections. In additional calculations (not depicted) we have varied $d_x\neq d$ while fixing the separations $d=d_y=d_z$. Indeed, the sign of the effective Hall voltage does {\it not} change when changing the sign of $d_x$. 

The overall reasoning is analogous for the Hall voltage due to the tori parallel to the $xz$-plane. Again, the tori parallel to the $yz$-plane pick up the local Hall voltage and guide it to the outside world.

In a strictly two-dimensional system parallel to the $xy$-plane, the crucial $yz$-tori do not exist. Thus, intuitively, the Hall-effect sign-inversion is not possible in two dimensions---as was previously proven mathematically \cite{Brian2009,Brian2012}.

Let us note in passing that our intuitive reasoning suggests that a related sign-inversion is expected for the Ettingshausen-Nernst effect. Herein, a temperature gradient along the $x$-direction (instead of a potential gradient for the Hall effect) leads to a heat current and to an electric current in the $x$-direction. Together with a magnetic field along $z$, in analogy to the Hall effect, this current leads to a transverse voltage in the $y$-direction. Again, depending on the tori separation $d$, i.e., on how this voltage is picked up internally, we expect a positive or negative external Ettingshausen-Nernst voltage---for the same constituent material.

Single-component structures like shown in Fig.\,4 could be fabricated by three-dimensional direct laser writing (3D DLW) \cite{Freyman2010} of a polymer scaffold, followed by replication using a double-inversion procedure for silicon \cite{Tetreault2006} or titania \cite{Frolich2013}. This technology has originally been developed for photonic crystal fabrication. Recently, 3D DLW has been enhanced by rapid galvo scanners \cite{Bueckmann2014}, such that structures with sub-micron feature sizes and, at the same time, overall volumes of some ${\rm mm}^3$ can be fabricated within some hours of writing time. However, double-inversion procedures are still quite demanding technologically. It is much simpler and more reliable to just conformally coat the polymer scaffolds by a semiconductor material. For example, promising results have been obtained by using atomic-layer deposition of ZnO doped with Al (\guil{AZO}) at variable concentrations \cite{Frolich2011,Luka2011,Dhakal2012}. The resulting geometry, however, is distinct from the one considered so far. 

We have thus repeated the calculations for hollow tori (illustrated in Fig.\,9(a)) instead of massive tori (see Fig.\,4). Hollow tori are a reasonable approximation because the polymer scaffold can be considered as an isolator compared to the semiconductor and is thus electrically equivalent to vacuum/air. For consistency, we continue using parameters for n-doped silicon as constituent material. Corresponding results are exhibited in Fig.\,9(b). Again, the overall qualitative behavior is qualitatively rather similar to what we have discussed above, once again highlighting the robustness of the sign-inversion effect for the single-constituent metamaterial. The moduli of the Hall voltages for $r/R=0.14$ are about six times larger for the hollow than for the solid tori because the current is restricted to a thinner layer along the $z$-direction. As discussed in the introduction, the Hall voltage scales inversely with the thickness in the $z$-direction.

In conclusion, we have shown that the Hall voltage of a suitably shaped three-dimensional porous material and that of an otherwise identical bulk material can have opposite sign. This means that, in effect, a hole conductor appears like an electron conductor or vice versa. This finding challenges the common textbook wisdom on the connection between the dominant type of charged carrier and the sign of the Hall voltage. We have argued that it should be possible to realize such metamaterial structures using state-of-the-art three-dimensional micro-fabrication technology. On the basis of our intuitive discussion of the calculation results, we expect a related sign-inversion for the Ettingshausen-Nernst effect on the same metamaterial structures.

We thank the Hector Fellow Academy and the Karlsruhe School of Optics \& Photonics (KSOP) for support.


%

%

\end{document}